\documentclass{webofc}
\usepackage[varg]{txfonts}   
\usepackage{listings}
\usepackage{lineno}
\usepackage{hyperref}
\begin{document}
%
\title{Coffea}
\subtitle{Columnar Object Framework For Effective Analysis}

\author{
  \firstname{Nicholas} \lastname{Smith}\inst{1}\fnsep\thanks{\email{nick.smith@cern.ch}}
  \and \firstname{Lindsey} \lastname{Gray}\inst{1}
  \and \firstname{Matteo} \lastname{Cremonesi}\inst{1}
  \and \firstname{Bo} \lastname{Jayatilaka}\inst{1}
  \and \firstname{Oliver} \lastname{Gutsche}\inst{1}
  \and \firstname{Allison} \lastname{Hall}\inst{1}
  \and \firstname{Kevin} \lastname{Pedro}\inst{1}
  \and \firstname{Maria} \lastname{Acosta}\inst{1}
  \and \firstname{Andrew} \lastname{Melo}\inst{2}
  \and \firstname{Stefano} \lastname{Belforte}\inst{3}
  \and \firstname{Jim} \lastname{Pivarski}\inst{4}
}

\institute{
  Fermi National Accelerator Laboratory
\and
  Vanderbilt University
\and
  INFN
\and
  Princeton University
}

\abstract{%
  The coffea framework provides a new approach to High-Energy Physics analysis, via columnar operations, that improves time-to-insight,
  scalability, portability, and reproducibility of analysis.
  It is implemented with the Python programming language, the scientific python package ecosystem, and commodity big data technologies.
  To achieve this suite of improvements across many use cases, coffea takes a factorized approach, separating the analysis implementation and data delivery scheme.
  All analysis operations are implemented using the NumPy or awkward-array packages which are wrapped to yield user code whose purpose is quickly intuited.
  Various data delivery schemes are wrapped into a common front-end which accepts user inputs and code, and returns user defined outputs.
  We will discuss our experience in implementing analysis of CMS data using the coffea framework along with a discussion of the user experience
  and future directions.
}
\maketitle

\section{Introduction}
\label{sec:intro}
The present challenge for High-Energy Particle Physics (HEP) data analysts is daunting: due to the
success of the Large Hadron Collider (LHC) data collection campaign over Run 2 (2015-2018),
the Compact Muon Solenoid (CMS) detector has amassed a dataset of order 10 billion proton-proton
collision events. The raw detector information is reconstructed into high-level information,
such as the trajectories of visible outgoing subatomic particles, using a centrally-maintained
software~\cite{CMSSW} and distributed computing system~\cite{WLCG}.
Even after significant processing and distillation, this high-level summary of each
collision event still contains order 1 kilobyte of compressed data~\cite{NanoAOD}. The CMS physicist/data-analyst
is tasked with processing the resulting tens of terabytes of distilled data (along with a similar
magnitude of simulation data) in a mostly autonomous fashion, typically designing (or inheriting) a processing framework
written in C++ or Python using a set of libraries known as the ROOT framework~\cite{ROOT}, and
parallelizing the processing over distributed computing resources using HTCondor~\cite{Condor}
or similar high-throughput computing systems. Each physicist is interested in a different subset of this data,
and will look at a different projection of the high-level variables in order to produce summary
statistics, such as histograms, from which statistical inference can be made to measure parameters
of, or probe compatibility with, theoretical expectations.

As the collected and simulated datasets grow, the development and maintenance burden of this analysis
code represents an increasingly large fraction of physicists' time. The scale of processing
requires physicists to be increasingly cognizant of the structure and performance characteristics
of their code, despite the fact that they are not traditionally trained as software engineers.
Simultaneously, physicists are more often utilizing modern machine learning techniques and their associated
libraries (e.g., Tensorflow, PyTorch), which enforce certain data structures that are atypical
in traditional approaches to analysis. Several hundred physicists face these challenges now, and
they will only magnify as the High-Luminosity LHC upgrade is projected to produce datasets over
an order of magnitude larger. Now is the time to investigate novel approaches to HEP data analysis
that are easy to use, scalable, and fast.

HEP physicists have been called upon to produce bespoke data analysis applications for several
decades, historically at the forefront of data volume and computational requirements. However, 
the needs of the private sector, and also of other scientific disciplines, have reached or
surpassed the scale of HEP data analysis. One of our core goals is to investigate the
applicability of solutions found outside HEP towards our data analysis needs.
In these proceedings, we introduce the concept of columnar analysis and the coffea
framework, then discuss the user experience and scalability characteristics of the framework,
and propose future directions for analysis systems research and development that we will pursue.

\section{Columnar Analysis}
\label{sec:columnar}
\begin{figure}[h]
  \centering
  \includegraphics[width=0.7\textwidth]{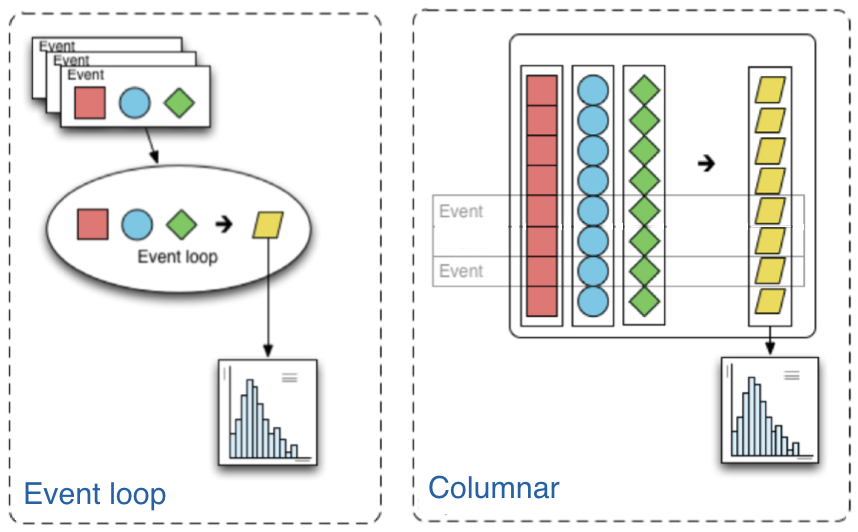}
  \caption{Schematic of the event-loop (left) and columnar (right) data processing paradigms.}
  \label{fig:columnar}
\end{figure}

Columnar analysis is a paradigm that describes the way the user writes the analysis application that is best described
in contrast to the the traditional paradigm in HEP of using an event loop.
In an event loop, the analysis application operates in a record-oriented manner on the input data,
where each reconstructed particle collision event record is deserialized and loaded into a
structure containing several fields, such as the properties of the visible outgoing particles
that were reconstructed in a collision event. The analysis code manipulates this structure to either output derived
quantities or summary statistics in the form of histograms. In contrast, columnar analysis operates
in a column-oriented manner, where individual columns represent portions of data spanning a \emph{chunk}
(i.e., partition, batch) of event records are manipulated using array programming primitives in turn, to compute
derived quantities and summary statistics. A visualization of the two contrasting approaches is
shown in Fig.~\ref{fig:columnar}.

Array programming is widely used within the scientific python ecosystem, supported at the
foundational level by the NumPy~\cite{NumPy} library, with functionality extended by libraries such
as SciPy~\cite{SciPy}, Pandas~\cite{Pandas}, and others.
However, although the existing scientific python stack is fully capable of analyzing rectangular
arrays (i.e., no variable-length array dimensions), HEP data is very irregular, and manipulating
it can become awkward without first generalizing array structure a bit. The
awkward-array~\cite{awkward} library performs this function, extending array programming capabilities
to the complexity of HEP data. In particular, awkward-array extends NumPy \emph{broadcasting} semantics to
so-called \emph{jagged} arrays, where the number of elements varies per sub-array.
In addition, awkward-array allows Structure-of-Array (SoA) objects to be
manipulated and viewed as though they were plain structures, so that intuitive expressions can be
written that access individual arrays of data.

\section{The coffea framework}
\label{sec:coffea}

The coffea framework is a python package that is indexed and installable~\footnote{
  \texttt{pip install coffea}, other options documented at \url{https://coffeateam.github.io/coffea/installation.html}
} via standard 
python packaging infrastructure, which provides several recipes and utilities to aid in development
of a HEP analysis within the columnar analysis paradigm.  The package has developed organically
to support our initial goal of implementing CMS data analyses within the
scientific python ecosystem. Here, we define ``analysis'' as the process of selecting events of
interest based on the high-level input variables or functions thereof, extracted from ROOT files or similar
column-oriented serialization formats, transforming numeric quantities, and computing summary statistics or low-volume tabulated
data for use with statistical inference tools. As this is the first full attempt at using scientific python
libraries for CMS analyses, certain missing extensions which normally would be provided by ROOT libraries or by
CMS experiment-specific libraries had to be ported to a columnar paradigm. Examples include:
\begin{itemize}
  \item multidimensional histogram objects that are serializable and mergeable, with support for
    both categorical and numeric axes, that are fillable by arrays;
  \item certain experiment-specific corrections that are applied to simulated data, typically as a
    piecewise function of some set of event parameters;
  \item utilities to reduce boilerplate necessary to construct the awkward arrays that represent
    the events SoA object; and
  \item wrappers to enable use of novel scale-out mechanisms, as discussed further in
    Section~\ref{sec:scalability}.
\end{itemize}

As many of these extensions are performance-critical, we make full use of the vectorized array
programming primitives provided by NumPy and SciPy. We also utilize Numba~\cite{Numba}, which
just-in-time compiles a subset of python and NumPy code into machine code, for operations where
no efficient composition of existing array programming primitives could be found. In general, any
operation performed along the critical dimension---namely, per event---is not performed sequentially
within the python interpreter but by vectorized machine code on the arrays forming the high-level
SoA objects.

\section{User experience}

All coffea core developers actively participate in CMS data analyses.  Along with the core developers,
several other data analysts have started adopting the coffea framework within their analysis workflows.
Some users have contributed to the package, extending coffea where their analysis required additional
functionality. As the package is open-source and available on GitHub~\footnote{\url{https://github.com/CoffeaTeam/coffea}}, any interested
party can submit patches to the source to add features or contribute bugfixes.

The main feedback from end users is that they appreciate being able to write analysis code in python while
still maintaining good performance, such that their workflows complete quickly. The user time and barrier
to entry for python is sufficiently low for physicist-programmers that having the entire workflow code
in python is a useful feature. In existing python bindings for ROOT, the lack of a vectorized infrastructure
along the event dimension leads to poor-performing workflows that inhibited the ability for users to develop
and test their analysis.\footnote{
  ROOT has also addressed this issue with the introduction of \texttt{RDataFrame}, which
  can help lower more analysis operations from python interpreted code into compiled machine code.
} Users who do not already have extensive experience with the ROOT ecosystem have reported that the
scientific python ecosystem has excellent ``google-ability'', i.e., example solutions for many tasks
are easily found and supporting documentation is quite extensive. By depending on the large
user and developer base of scientific python, the coffea framework requires little developer effort
to maintain user support, since the underlying infrastructure often provides the necessary support.

In some cases, no efficient composition of array programming primitives can provide the end user with
the necessary solution for a given task. This happens when the algorithm to be implemented, or intra-event
data on which it operates, becomes sufficiently complex. In these cases, users must learn how to write their own array
programming kernels, e.g., with Numba. Presently, these kernels can be challenging to write. Work is ongoing
within the awkward-array library to simplify such developments. For some widely-used algorithms, user-developers
have contributed the missing algorithms to the coffea codebase.

One important benefit of utilizing the scientific python ecosystem is the seamless integration with
Jupyter notebooks~\cite{Jupyter}. Jupyter notebooks combine documentation, source code, and partial results into one persistent
document that provides much faster iterative data exploration than the traditional command-line interface.
Notebooks can be effective as teaching aids, and may enable analysis preservation efforts.
However, users still find the traditional approach valuable for established workflows.

\section{Scalability}
\label{sec:scalability}

One of the challenges for HEP data analysts is \emph{scale-out}: the process of scaling their workflow from
a small test sample to tens of terabytes. Luckily, essentially all HEP workflows are trivially parallel as each event
is statistically independent. Analysts build their own map-reduce algorithm to divide some set of input
files into a set of batch jobs, and reduce the job output data by adding bin counts in histograms.
With coffea, we provide a high-level wrapper around user analysis code: the coffea \emph{processor}.
The \texttt{ProcessorABC} abstract class defines an interface where the user receives a chunk of
columns and is responsible for returning a reducible object, where reducibility is enforced by
the \texttt{AccumulatorABC} abstract class. The most common reducible object is a histogram.

%
%
%
%
%
%

With the processor defined, the responsibility of dividing the work over a given execution resource
and accumulating the results is taken by the coffea framework. The currently supported coffea
executors are:
\begin{enumerate}
  \item
    a local executor utilizing parallel python processes;
  \item
    an execution wrapper for the Apache Spark~\cite{Spark} distributed computing platform;
  \item
    an execution wrapper for the Dask~\cite{Dask} distributed computing platform; and
  \item
    an execution wrapper for the Parsl~\cite{Parsl} distributed computing platform.
\end{enumerate}
The user can then run their processor on a small set of test input files or on a full dataset
by simply changing the executor method within the coffea framework.

\section{Future directions}

Our experience with the coffea executors highlights the need for intermediate-scale resources:
a $\lesssim 100$ core-hour resource is sufficient to process an entire CMS analysis over run 2 data in
minutes with the appropriate executor. Further speed-up will only happen with the introduction of
derived column caches in a multi-user context. Therefore, we are now investigating
the concept of analysis facilities, dubbed \emph{coffea farms}, where several users operate on the same
datasets, and the framework ensures that users accessing the same data will access a cached copy of it.
It is envisioned that several small clusters will host coffea farm applications
for small groups of CMS data analysts, enabling rapid analysis prototyping with large-scale datasets,
and removing manual work to reduce the input datasets for efficient re-analysis
(often referred to as \emph{slimming and skimming}).

\section{Conclusions}
Columnar analysis is an effective paradigm for HEP data analysis within the CMS collaboration,
and we have implemented several maturing analyses in a columnar fashion. We have profited from the scientific
python ecosystem to provide a library that is largely feature-complete for our needs with minimal direct
developer resources. The coffea framework enables users to write high level operations, while the library
code maintains performance. Coffea simplifies the interface to novel scale-out mechanisms, and provides
an avenue to investigate the resource gains that may be found with a multi-user common-framework analysis
facility.


\end{document}